\documentclass[a4paper,11pt]{article}
\usepackage{pos}
\usepackage{float}
\usepackage{subcaption}
\usepackage{multicol}
\usepackage{enumitem}

\title{\begin{flushright}
\small{
WUB/22-04\\}
\vskip 0.7cm
\end{flushright}
Optimized meson operators for charmonium spectroscopy and mixing with glueballs}
\ShortTitle{Optimized meson operators for charmonium spectroscopy and mixing with glueballs}

\author*[a]{Juan Andrés Urrea-Niño}
\author[a]{Francesco Knechtli}
\author[a]{Tomasz Korzec}
\author[b]{Michael Peardon}

\affiliation[a]{Dept. of Physics, Bergische Universit{\"a}t Wuppertal,\\
Gau{\ss}strasse 20, 42119 Wuppertal, Germany}
\affiliation[b]{School of Mathematics, Trinity College,\\
Dublin 2, Ireland}

\emailAdd{urreanino@uni-wuppertal.de, knechtli@uni-wuppertal.de, korzec@uni-wuppertal.de, mjp@maths.tcd.ie}

\abstract{Optimized meson operators in the distillation framework are used to study the charmonium spectrum in two ensembles with two heavy dynamical quarks at half the physical charm quark mass but different lattice spacings. The use of optimal meson distillation profiles is shown to increase the overlap with the ground state significantly, as well as grant access to excited states, for multiple quantum numbers including hybrid states with very little additional cost. These same operators are also employed for the calculation of meson-glueball mixing.}

\FullConference{%
  The 39th International Symposium on Lattice Field Theory (Lattice2022),\\
  8-13 August, 2022 \\
  Bonn, Germany 
}

\begin{document}
\maketitle

\section{Introduction}
The smearing method known as distillation \cite{Peardon} is often used in hadron spectroscopy due the numerous advantages it has. In the case of mesons it has been been shown to give clear access to different quantum numbers $J^{PC}$ of interest via local and derivative-based operators. But it also does so with a fixed, although considerable, computational cost \cite{Peardon,Liu,Dudek2}. This cost mainly comes from solving multiple linear systems with the Dirac operator and a number $N_v$ of eigenvectors of the 3D covariant Laplace operator corresponding to the eigenvalues with smallest absolute value. The solutions to these linear systems are used to build low-dimensional all-to-all propagators known as \textit{perambulators} which can be stored and used multiple times since they are independent from the operators that define the different symmetry channels. The different operators that can be used give rise to the \textit{elementals}, also low-dimensional matrices, which can then be appropriately combined with the perambulators to calculate the relevant correlation functions. The explicit entries of the perambulators are given by
\begin{equation}
\tau[t_1,t_2]^{ij}_{\alpha \beta} = v_{i,\alpha}[t_1]^{\dagger} D^{-1} v_{j,\beta}[t_2],
\end{equation}
where $D$ is the Dirac operator and $v_{i,\alpha}[t_1]$ corresponds to a vector which contains the $i$-th Laplacian eigenvector at time $t_1$ in Dirac index $\alpha$ and is zero everywhere else. The entries of the elementals are given by
\begin{equation}
\Phi[t]^{ij}_{\alpha \beta} = v_{i,\alpha}[t]^{\dagger} \Gamma v_{j,\beta}[t],
\end{equation}
where $\Gamma$ is the operator which defines the symmetry channel. The computational work for the construction of the perambulators corresponds to the solution of $4\times N_t \times N_v$ for a single gauge configuration with $N_t$ the temporal extent of the lattice. Additionally, for a fixed level of smearing the value of $N_v$ scales with the 3D physical volume of the lattice \cite{Peardon}. Therefore, for sufficiently large lattices and large statistics the number of inversions is considerably high. At light quark masses close to their physical values each inversion is also expensive due to worse conditioning, which adds up to the total cost. All of these considerations point to the need to address these costs. A guiding question is how to choose $N_v$ since too small would neglect significant low energy modes while too big would increase the overall cost too much and include non-significant high energy modes. Not only this, but it is also important to know if a given choice of $N_v$ is equally useful for different operators of interest and if all the used eigenvectors will contribute in the same manner. This is not expected a priori since local and derivative-based operators sample different spatial structures which can also differ in between different excitations of a same channel. An initial study of the use of distillation profiles to not only build operators that optimally use each available eigenvector but also to qualitatively determine an appropriate choice of $N_v$ for each operator and energy level studied was presented in \cite{Urrea}. Here an extension of this work is presented including the application of the proposed method to analyze multiple meson operators corresponding to different $J^{PC}$ using two $N_f = 2$ ensembles with same quark mass but different lattice spacings and volumes to check the effectiveness of the method and the obtained so-called optimal meson distillation profiles with the volume scaling, as well as the mixing of these optimized meson operators with glueball operators.  

\section{Construction of optimal meson distillation profiles}
As described in \cite{Urrea}, a basis of $N_B$ different quark gaussian profiles $g_k(\lambda)$ are used to define distilled quark fields which are then used to build a basis of $7$ meson operators $\mathcal{O}_k$ with a fixed $\Gamma$ and projected to zero spatial momentum. From these meson operators a $7\times 7$ correlation matrix is built as
\begin{equation}
C_{ab}(t) = \left \langle \mathcal{O}_a(t) \bar{\mathcal{O}}_b(0)\right \rangle
\end{equation}
which is then pruned via a singular value decomposition for numerical stability as described in \cite{Balog,Niedermayer} to keep the most significant $N_S = 4$ operators. It should be noted that since the operator $\Gamma$ is fixed this matrix can be built at no significant extra cost since the only necessary step is to replace the profile factor $g_k(\lambda_i[t])^{*}g_k(\lambda_j[t])$ that multiplies entry $\Phi[t]^{ij}_{\alpha \beta}$ of the corresponding elemental. The resulting $N_S\times N_S$ pruned $C(t)$ is then used to set up the GEVP formulation \cite{Michael,Luscher,Blossier} as
\begin{equation}
C(t) u_e(t,t_G) = \rho_{e}(t,t_G) C(t_G) u_e(t,t_G),
\end{equation} 
where $u_e(t,t_G)$ are the generalized eigenvectors and $\rho_{e}(t,t_G)$ their corresponding generalized eigenvalues with $e=0,...,N_S - 1$ ordered such that $\rho_{e}(t,t_G) > \rho_{e+1}(t,t_G)$. From $\rho_{e}(t,t_G)$ one can extract the effective mass of energy level $e$ while from $u_{e}(t,t_G)$ one can extract the coefficients that define a linear combination of the pruned operators which has the largest overlap with the actual energy eigenstate of the corresponding $J^{PC}$. As shown in \cite{Urrea}, the elemental corresponding to this optimal operator for a fixed $\Gamma$ and energy level $e$ is given by
\begin{equation}
\tilde{\Phi}[t]^{ij}_{\alpha \beta} = \tilde{f}^{(\Gamma,e)}(\lambda_i[t],\lambda_j[t]) v_{i,\alpha}[t]^{\dagger} \Gamma v_{j,\beta}[t], 
\end{equation}
where the optimal meson distillation profile $\tilde{f}^{(\Gamma,e)}(\lambda_i[t],\lambda_j[t])$ is given by
\begin{equation}
\tilde{f}^{(\Gamma,e)}(\lambda_i[t],\lambda_j[t]) = \sum_{k} \eta^{(\Gamma,e)}_k g_k(\lambda_i[t])^{*} g_{k}(\lambda_j[t])
\end{equation}
and the coefficients $\eta^{(\Gamma,e)}_k$ take into account the coefficients from the generalized eigenvector $u_e(t,t_0)$ and also the singular vectors from the pruning. This profile not only determines how the vectors $v_{i,\alpha}[t]$ and $v_{j,\beta}[t]$ must be weighted in the elemental for a fixed $\Gamma$ and energy level $e$ but also shows if a sufficient level of suppression of high Laplacian eigenvalues has occurred and the chosen value of $N_v$ can be considered acceptable.

\section{Meson results in $N_f = 2$ QCD}
The model used in this work corresponds to $N_f = 2$ QCD clover-improved Wilson fermions with quark mass at half of the physical charm quark mass. Two different ensembles are used, one with size $48\times 24^3$ and lattice spacing $a\approx 0.0658$ fm \cite{Urrea} and the other one with size $96\times 48^3$ and lattice spacing $a\approx 0.049$ fm \cite{Cali}, both with periodic boundary conditions in time for the gauge links. The coarsest lattice, which also has the smallest 3D physical volume, is the starting point of the analysis with $N_v = 200$. A basis of $7$ different quark gaussian profiles
\begin{equation}
g_k(\lambda) = e^{-\frac{\lambda^2}{2\sigma_k}},
\end{equation}
with widths chosen to allow different ranges of suppression for the relevant eigenvalues and whose specific values can be found in \cite{Urrea}, are used to build the meson operators. For the finer lattice these widths are scaled according to the squared lattice spacing and a value of $N_v = 325$ is used, which via the volume scaling corresponds roughly to $100$ eigenvectors in the coarser lattice. Both local and derivative-based $\Gamma$ operators shown in \cite{Urrea}, the latter taken from \cite{Dudek}, are analyzed in both available ensembles for the iso-vector channel and the effective masses of the different $J^{PC}$ channels are extracted from the eigenvalues of the previously described GEVP formulation using $t_G=3$ and pruning at this same value of time. Fig. \ref{fig:Masses_Em1} shows the effective masses for a selection of operators using the optimal profiles, standard distillation and stochastic estimation without any smearing (only for the local operators) for the sake of comparing the three methods in the coarsest lattice.  

\begin{figure}[H]
\centering
\begin{subfigure}[b]{0.4\linewidth}
\includegraphics[width=\linewidth]{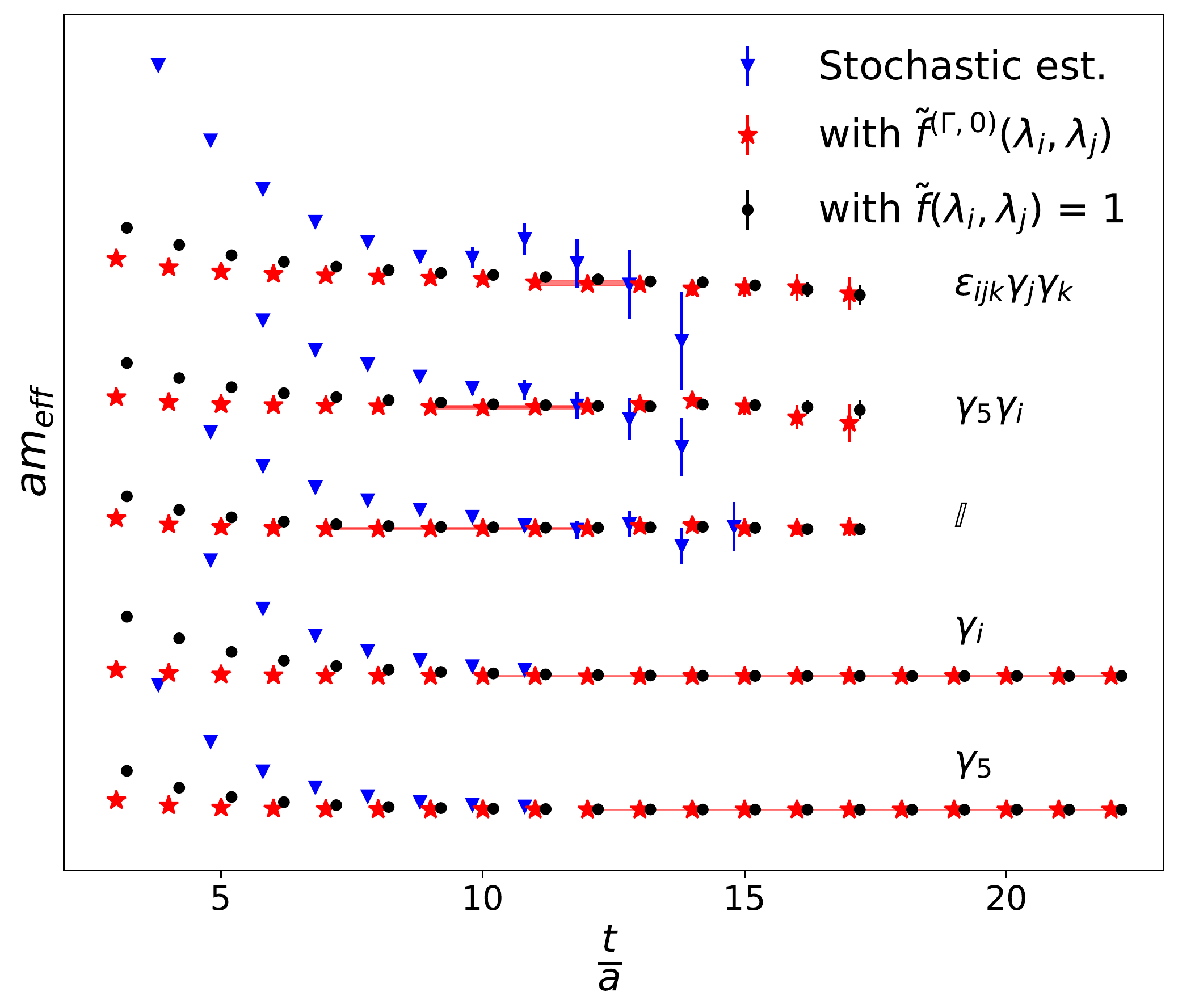}
\caption{Masses of local operators using optimal profiles, standard distillation and stochastic estimation. Masses are displaced for clarity.}
\end{subfigure}
\begin{subfigure}[b]{0.4\linewidth}
\includegraphics[width=\linewidth]{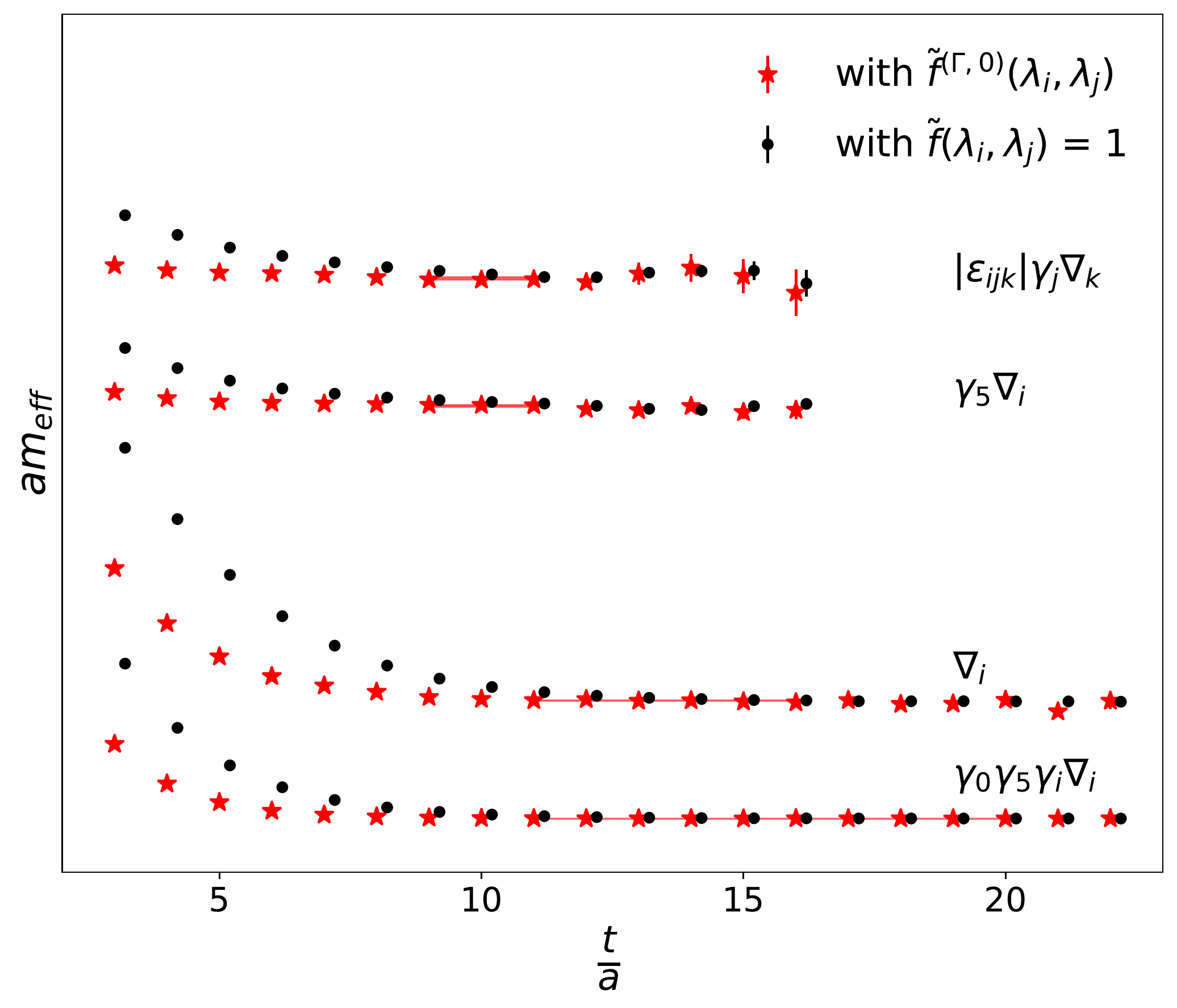}
\caption{Masses of derivative-based operators using optimal profiles and standard distillation. Masses are displaced for clarity.}
\end{subfigure}
\caption{Masses of a selection of operators in the coarse ensemble.}
\label{fig:Masses_Em1}
\end{figure}

It is clear that the use of the optimal meson distillation profiles leads to significant suppression of the excited-state contamination which in turn leads to earlier and in general longer mass plateaus when compared to standard distillation. This improvement can be numerically quantified via the so-called fractional overlap defined in \cite{Urrea}, a quantity which measures the presence of excited state contamination at early times which can be calculated from the correlation functions of each operators using both variants of distillation, where a value closer to $1$ means a larger suppression of the undesired contamination. The fractional overlaps for some of the analyzed operators are the following:
\vspace{-1pc}
\begin{multicols}{2}
\begin{itemize}
\item $\Gamma = \gamma_5$: 0.9272(3) $\rightarrow$ 0.9858(2)
\item $\Gamma = \gamma_i$: 0.8743(10) $\rightarrow$ 0.9900(5)
\item $\Gamma = \epsilon_{ijk}\gamma_j \gamma_k$: 0.77(7) $\rightarrow$ 0.93(1)
\item $\Gamma = \nabla_i$: 0.4758(7) $\rightarrow$ 0.742(2)
\item $\Gamma = \gamma_5 \nabla_i$: 0.84(1) $\rightarrow$ 0.970(5)
\item $\Gamma = \mathbb{Q}_{ijk} \gamma_j \nabla_k$: 0.858(8) $\rightarrow$ 0.981(3)
\end{itemize}
\end{multicols}

\begin{figure}[H]
\centering
\begin{subfigure}[b]{0.4\linewidth}
\includegraphics[width=\linewidth]{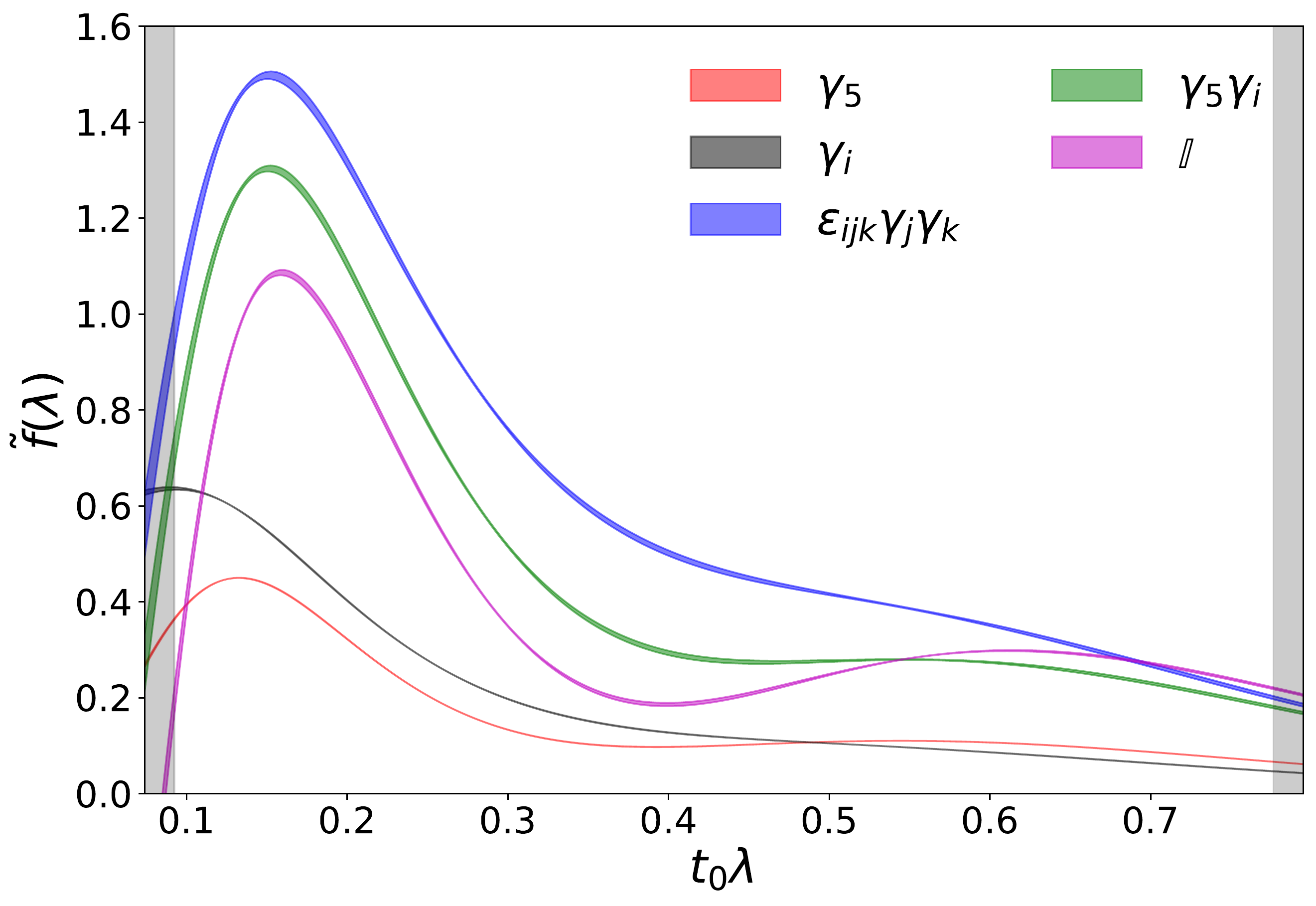}
\caption{Optimal meson distillation profiles of the ground state of the local $\Gamma$ operators as a function of the Laplacian eigenvalue.}
\label{fig:LocalProfsEm1}
\end{subfigure}
\begin{subfigure}[b]{0.4\linewidth}
\centering
\includegraphics[width=\linewidth]{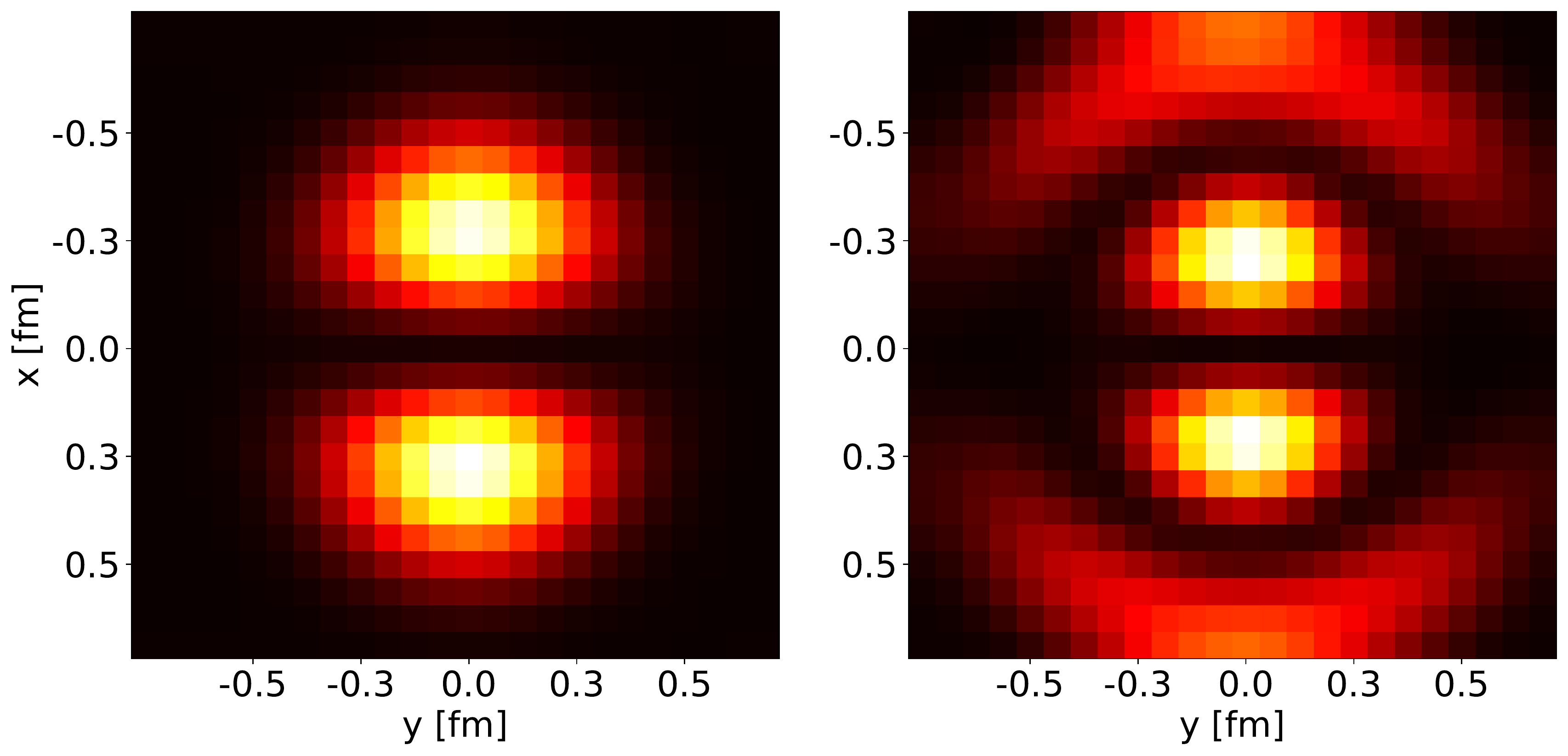}
\caption{Spatial profile of the ground and first excited states of the $\Gamma = \gamma_5 \nabla_1$ operator as defined in Eq. \ref{eqn:SpatialProfile} using the optimal meson distillation profile.}
\label{fig:SpatialProfile}
\end{subfigure}
\caption{Profiles in distillation and coordinate space of some of the analyzed operators.}
\label{fig:Masses_Em1}
\end{figure}

The significant closeness to $1$ of the fractional overlaps when the optimal profiles are used in this sample of values serves as further evidence of the advantage of using them. Given this improvement it is of interest to directly visualize the different profiles that for each operator and energy level yield the best overlap with the actual energy eigenstate. Their specific construction is given in \cite{Urrea} and for the case of the local operators in the coarse lattice they can be seen in Fig. \ref{fig:LocalProfsEm1} as a function of a single eigenvalue, made dimensionless by multiplying it with the scale $t_0$ \cite{LuscherScale}. Unlike the case of standard distillation none of them is a constant, which already points to the fact that a one-for-all approach with the profile is not the optimal alternative and different operators should be treated differently. Nonetheless the common suppression of larger eigenvalues shows that the intuition behind distillation of favoring small eigenvalues still remains valid. To check that this feature is not a result of basis bias, due to all the Gaussian quark profiles exhibiting this pattern, the GEVP formulation was also tried with a basis of monomials of the form $\lambda^{k}$ and the same optimal meson profiles were obtained. Additionally, the values of the profile at the largest available eigenvalues serves as a qualitative guide to determine if the chosen $N_v$ is large enough. Namely, if the profile has not decreased enough compared to its peak then more eigenvectors should be considered. It can be argued from Fig. \ref{fig:LocalProfsEm1} that for all local operators the chosen value of $N_v$ is large enough. As was also presented in \cite{Urrea} it is possible to visualize the spatial profile of the meson operator built using the corresponding optimal meson distillation profile, which for the case of $\Gamma =\gamma_5 \nabla_1$ is given by
\begin{equation}
\Psi^{(\gamma_5 \nabla_1, e)}(\vec{x}) = \frac{1}{N_t} \sum_{t=0}^{N_t - 1} ||Tr \left( \gamma_5 V[t]\tilde{\Phi}^{(\gamma_5 \nabla_1,e)}[t]V[t]^{\dagger} \right) \phi_0||_2^{2},
\label{eqn:SpatialProfile}
\end{equation}
where $e$ denotes the energy state, the norm is taken in color space, the trace is taken in Dirac space and $\phi_0$ is a 3D point source. The resulting spatial profile can be seen in Fig. \ref{fig:SpatialProfile}. This serves not only as a useful visualization tool to check the expected spatial behavior of the meson operators, e.g a P-wave structure for the $\Gamma = \gamma_5 \nabla_1$ operator with $1^{+-}$ numbers, but also to monitor finite-volume effects in cases when the extent of the profile is close to reaching the boundaries of the lattice. 

\begin{figure}[H]
\centering
\begin{subfigure}[b]{0.4\linewidth}
\includegraphics[width=\linewidth]{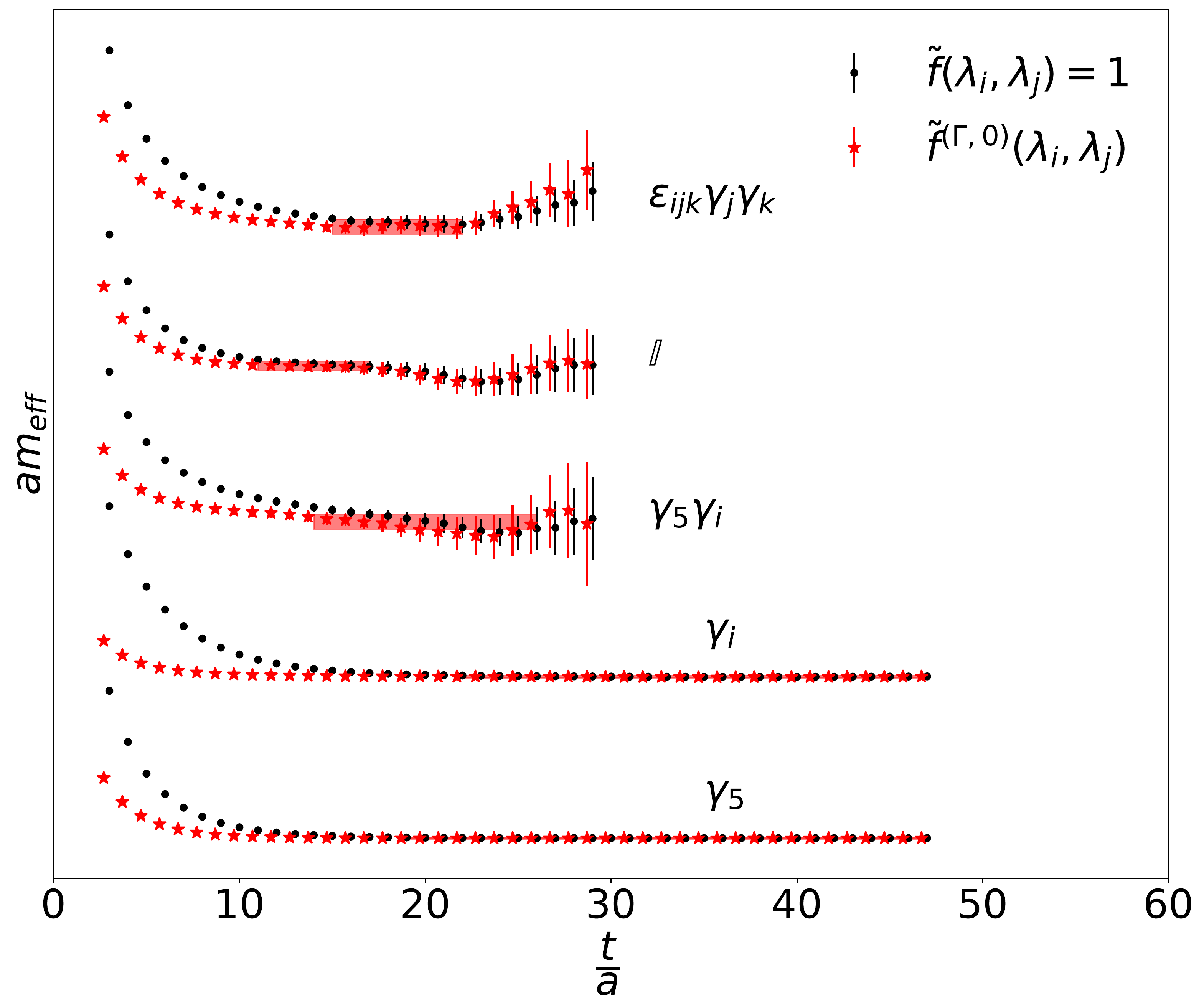}
\caption{Masses of local operators using optimal profiles, standard distillation and stochastic estimation. Masses are displaced for clarity.}
\end{subfigure}
\begin{subfigure}[b]{0.4\linewidth}
\includegraphics[width=\linewidth]{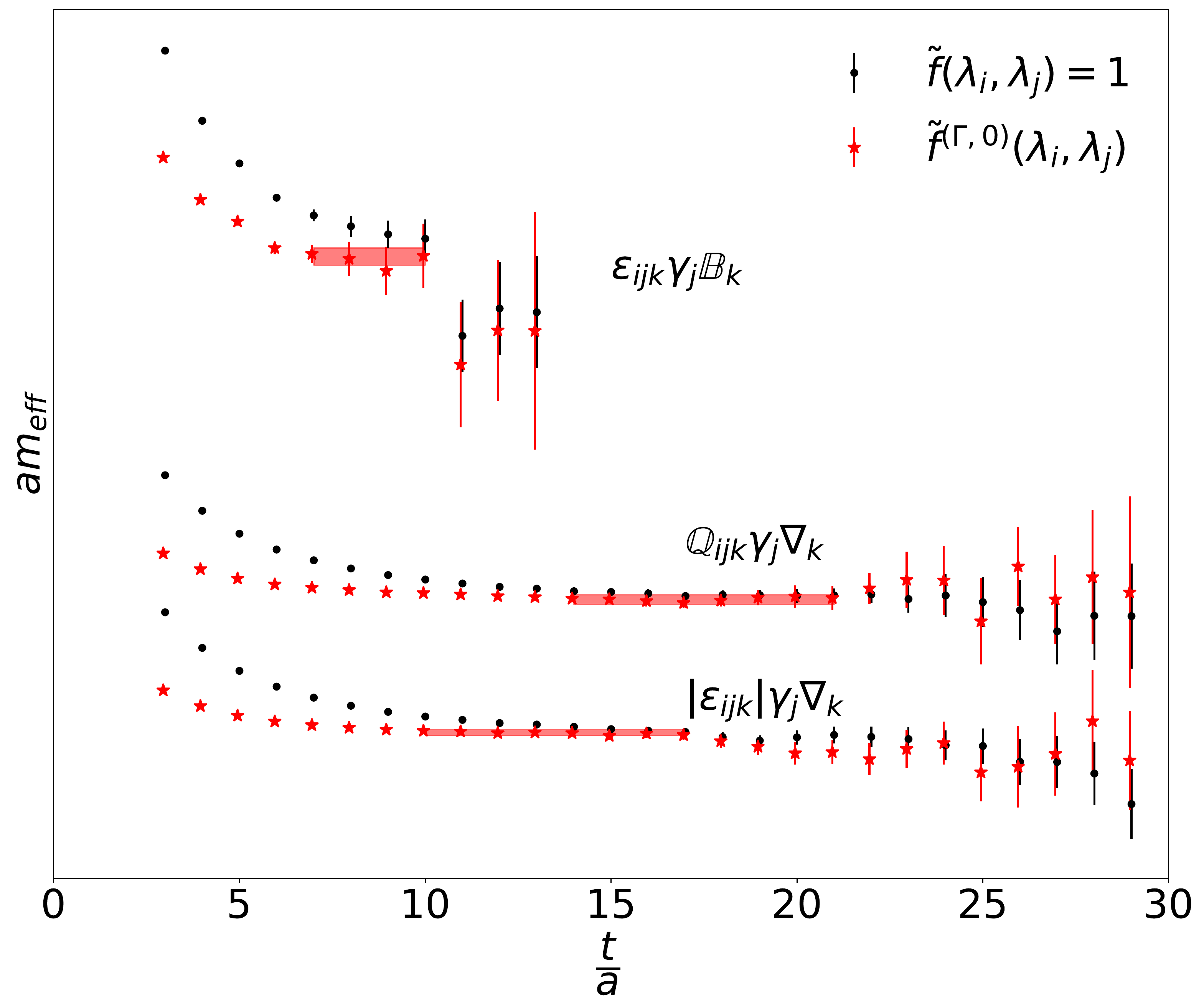}
\caption{Masses of derivative-based operators using optimal profiles and standard distillation. Masses are displaced for clarity.}
\end{subfigure}
\caption{Masses of a selection of operators in the fine ensemble.}
\label{fig:Masses_Nm1}
\end{figure}

\begin{figure}[H]
\centering
\begin{subfigure}[b]{0.4\linewidth}
\includegraphics[width=\linewidth]{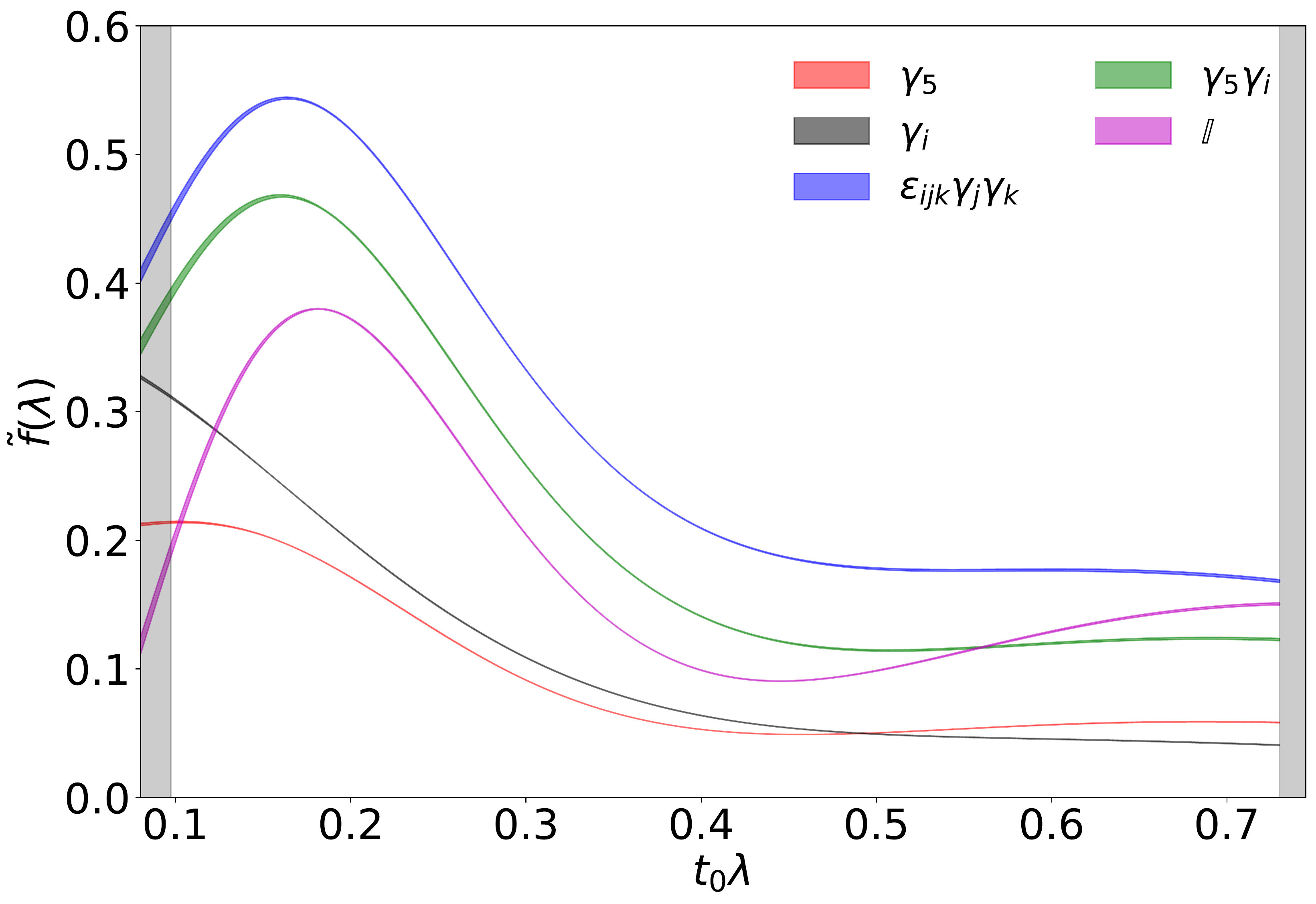}
\caption{Optimal meson distillation profiles of the ground state of the local $\Gamma$ operators as a function of the Laplacian eigenvalue.}
\label{fig:LocalProfsNm1}
\end{subfigure}
\begin{subfigure}[b]{0.4\linewidth}
\centering
\includegraphics[width=\linewidth]{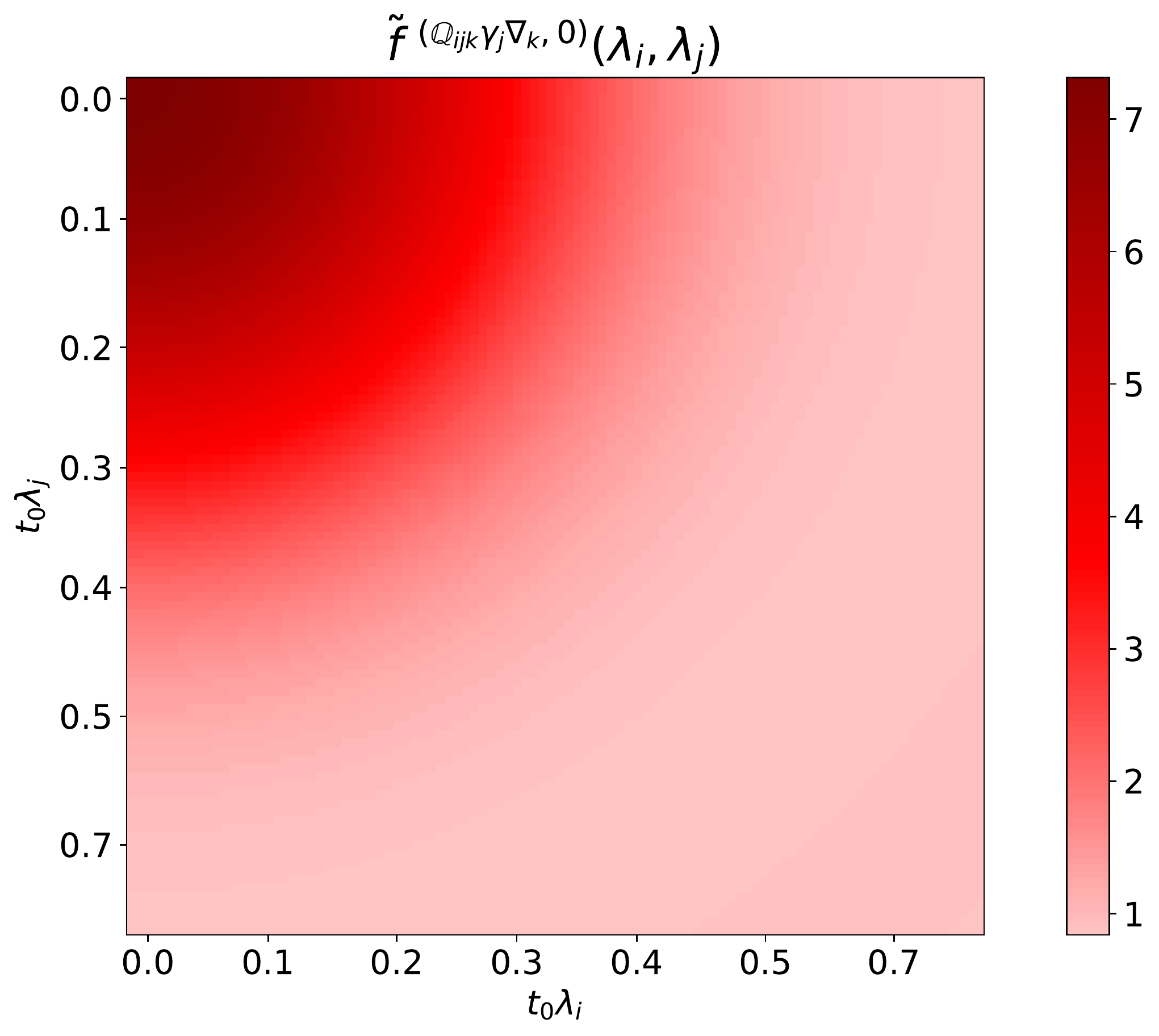}
\caption{Optimal meson distillation profile for the ground state of the $\Gamma = \mathbb{Q}_{ijk}\gamma_j \nabla_k$ operator as a function of two Laplacian eigenvalues.}
\label{fig:DerivProfNm1}
\end{subfigure}
\caption{Profiles in distillation space for some local and derivative-based operators in the fine ensemble.}
\label{fig:Profiles_Nm1}
\end{figure}

The same calculations can be performed for the ensemble with the finer lattice spacing. The corresponding results for the effective masses of some local and derivative-based operators can be seen in Fig. \ref{fig:Masses_Nm1}. Significant suppression of excited state contamination is again evidenced for both kinds of operators when the corresponding optimal meson distillation profiles are used. Some of the fractional overlaps for the studied operators are
\begin{multicols}{2}
\begin{itemize}
\item $\Gamma = \gamma_5$: 0.8765(7) $\rightarrow$ 0.9555(5)
\item $\Gamma = \gamma_i$: 0.825(3) $\rightarrow$ 0.969(2)
\item $\Gamma = \mathbb{Q}_{ijk}\gamma_j \nabla_k$: 0.82(2) $\rightarrow$ 0.92(1)
\item $\Gamma = \epsilon_{ijk} \gamma_j \mathbb{B}_k$: - $\rightarrow$ 0.91(1)
\end{itemize}
\end{multicols}
\vspace{-0.5pc}
where for the exotic $1^{-+}$ channel the hybrid operator $\epsilon_{ijk} \gamma_j \mathbb{B}_k$ only presents an effective mass plateau when the optimal profile is used. The distillation profiles can also be visualized for this ensemble, which is displayed in Fig. \ref{fig:Profiles_Nm1} for the ground state of the local operators as a function of a single eigenvalue and for a derivative based operator as a function of two eigenvalues. As mentioned before the widths of the Gaussian quark profiles involved are scaled appropriately and the displayed interval between the two gray regions corresponds to roughly 100 eigenvalues of the coarse lattice. All the observations made for the case of the coarse ensemble hold for these resulting profiles as the clearly different profiles distinguish the different channels and display the common suppression of higher eigenvalues. It is worth noticing that the overall shapes of the different profiles are similar when plotted against the dimensionless combination $t_0 \lambda$.

\section{Meson-glueball mixing in $N_f = 2$ QCD}
\begin{figure}[H]
\centering
\includegraphics[width=0.6\textwidth]{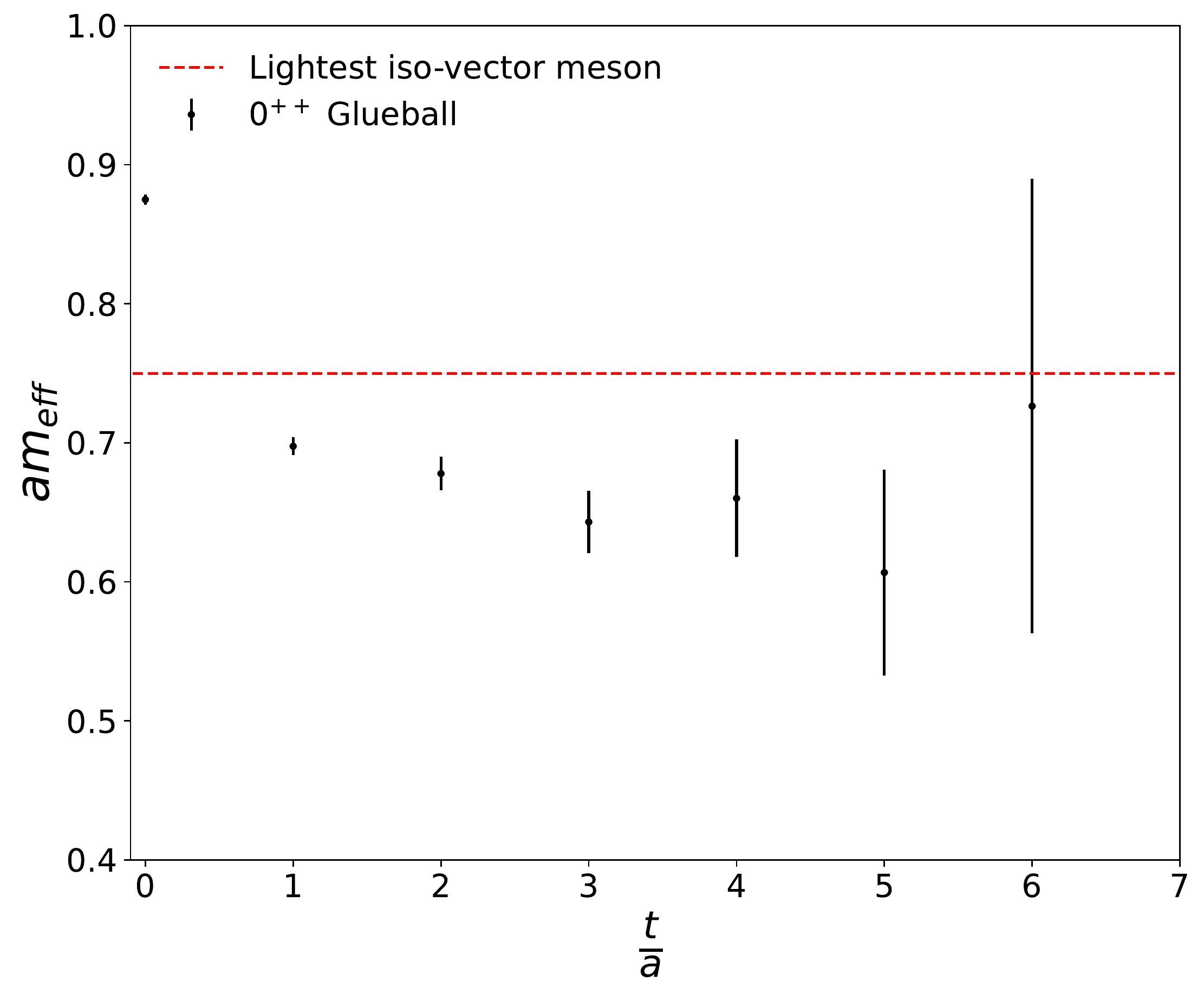}
\caption{Effective mass of the $0^{++}$ glueball in the coarse ensemble compared to the plateau average of the lightest iso-vector meson measured in \cite{Urrea}. The errorbars of the latter are omitted since they would not be visible at the scale of the plot.}
\label{fig:ScalarGlueball}
\end{figure}

\begin{figure}[H]
\centering
\begin{subfigure}[b]{0.4\linewidth}
\includegraphics[width=\linewidth]{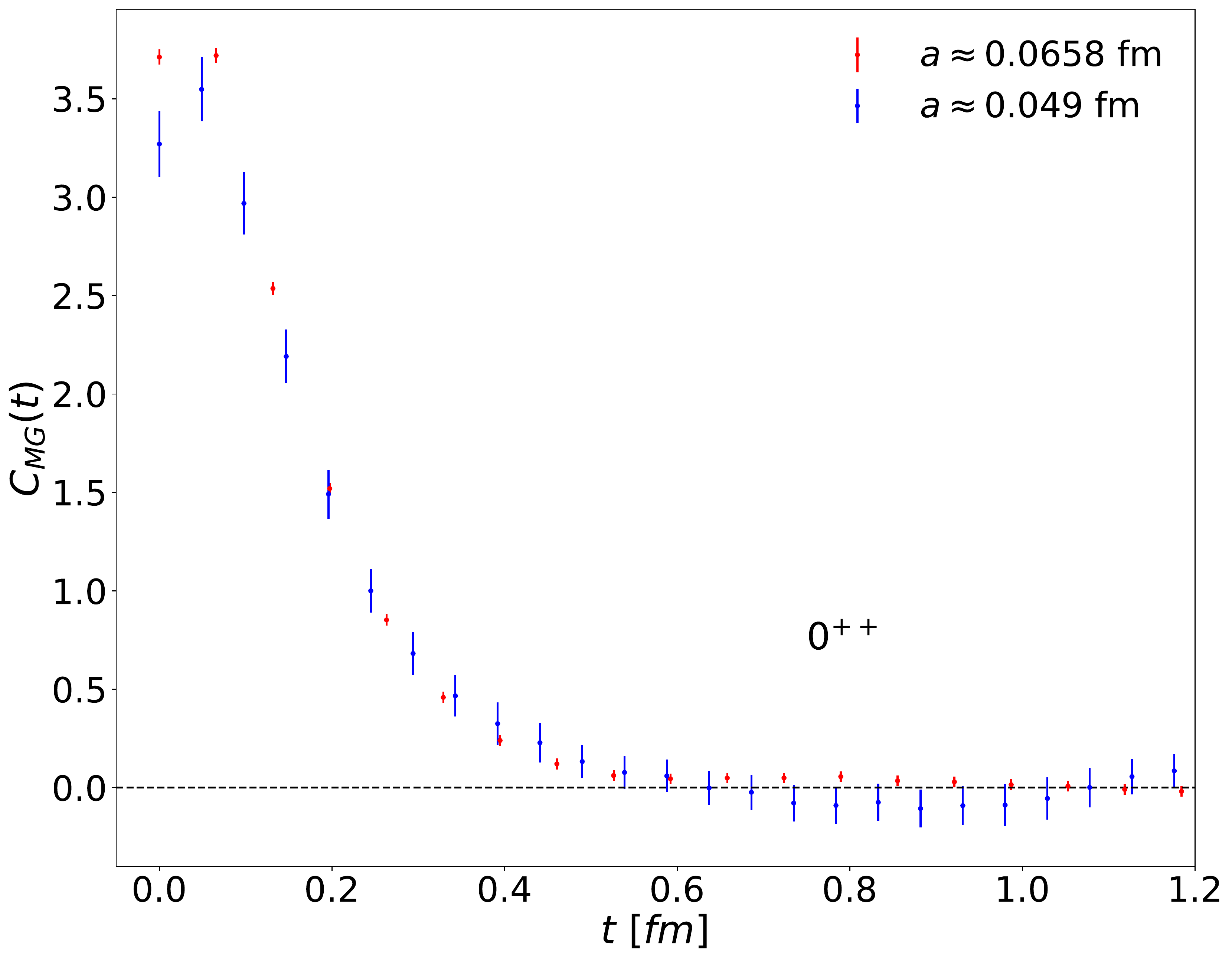}
\caption{Normalized mixing correlation for the $0^{++}$ channel in both ensembles.}
\end{subfigure}
\begin{subfigure}[b]{0.4\linewidth}
\centering
\includegraphics[width=\linewidth]{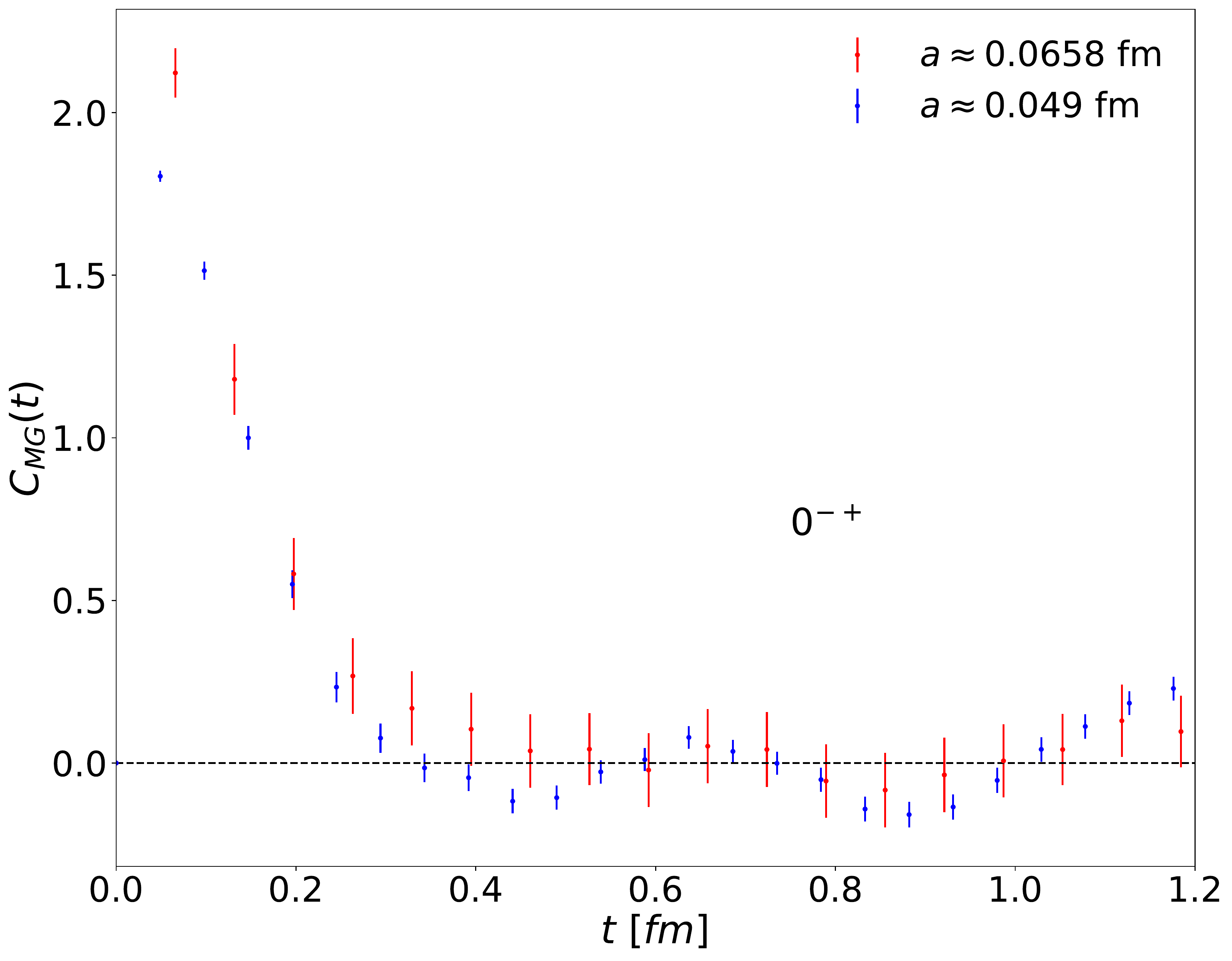}
\caption{Normalized mixing correlation for the $0^{-+}$ channel in both ensembles.}
\end{subfigure}
\caption{Normalized mixing correlations for the channels of interest in this work.}
\label{fig:Mixing}
\end{figure}

Iso-scalar meson operators can also be studied with the use of the meson profiles and a case of particular interest is their mixing with the corresponding glueball of the same symmetry channel. Since these iso-scalar meson operators require the inclusion of disconnected pieces in ther correlations, which tend to be lost to noise at very early values of times, the improvement brought by the meson profiles in earlier mass plateaus is specially desirable. For the case of the glueball operators, which are notoriously noisy, it is also necessary to work with the best possible operators. For this end the GEVP formulation is also adopted, where the operator basis is given by operators built from 3D Wilson loops with different shapes, windings and lengths \cite{PeardonGlueballs, Berg} involving link variables smeared via different smearing schemes, namely 3D HYP \cite{HYP} and 3D improved APE \cite{ImprovedAPE}. The $J^{PC}$ of interest for the mixing in this work are $0^{-+}$ and $0^{++}$ and therefore the Wilson loop operators have to be projected to the irreps $A_1^{-+}$ and $A_1^{++}$ accordingly \cite{Berg}. The correlation function corresponding to this meson-glueball mixing is given by
\begin{equation}
C_{MG}(t) = Tr\left( \Phi^{(\Gamma)}[t] \tau[t,t] \right) G^{(R^{PC})}(0),
\end{equation}
where $G^{(R^{PC})}(t)$ stands for the glueball operator in irrep $R^{PC}$ at time $t$ built using the GEVP eigenvectors. As a remark, since the $0^{++}$ glueball is the lightest particle in this study, as shown in Fig. \ref{fig:ScalarGlueball} for the coarse lattice, one would expect the clearest mixing signal to be in this channel. The results obtained for the mixing correlation functions for both ensembles are displayed in Fig.  \ref{fig:Mixing} for $0^{-+}$ and $0^{++}$, where each correlation function is normalized by dividing it by its value at a fixed time in physical units ($t_c = 0.245$ fm for $0^{++}$ and $t_c = 0.147$ fm for $0^{-+}$). This requires an interpolation of one of the correlation functions, in this case of the coarse lattice, which is performed in the regime where it behaves like an exponential decrease. For both symmetry channels and both ensembles there is a clear presence of a signal at early times which serves as evidence of the mixing. The meson operator used for the $0^{++}$ channel is $\Gamma = \mathbb{I}$ with standard distillation while for the $0^{-+}$ it is $\Gamma = \gamma_5$ using the optimal profile obtained from the iso-vector calculation. As expected, the $0^{++}$ displays the clearest signal out of the two channels analyzed. It should be noted that in both cases the noise of the correlation function is dominated by the noise of the glueball operators, which require significantly higher statistics than the meson operators to obtain a signal-to-noise ratio comparable to the latter. However, due to the fact that the meson operators are much more computationally expensive to calculate than the glueball ones it is the former the ones that dictate the statistics available. 

\section{Conclusions}
In this work an extension to the study of optimal meson distillation profiles presented in \cite{Urrea} was performed to compare two ensembles with $N_f = 2$ degenerate quarks at half the physical charm quark mass but with different lattice spacing and physical volume. In both ensembles the use of these profiles leads to a significant suppression of excited state contamination at no additional inversion cost, a main advantage of this modification. The construction of these meson profiles comes from a GEVP formulation where the previously unexploited choice of the quark distillation profile is used as an additional degree of freedom. Furthermore, these profiles help characterize optimal operators for different $J^{PC}$ of interest not only in distillation space but also in coordinate space via the visualization of their spatial profile. This characterization was performed for the case of meson operators, however it can be extended to general hadron operators where distillation is applicable as well as to the framework of stochastic distillation. Furthermore we did a first investigation of the mixing between isoscalar mesons and glueballs. A GEVP study of meson and glueball operators involving this mixing is an ongoing work, together with the application of the optimal meson distillation profiles in an $N_f = 3 + 1$ ensemble with a physical charm quark and three degenerate light quarks with the average mass as in nature.\\ \\

\noindent \textbf{Acknowledgement.} The authors gratefully acknowledge the Gauss Centre for Supercomputing e.V. (www.gausscentre.eu) for funding this project by providing computing time on the GCS Supercomputer
SuperMUC-NG at Leibniz Supercomputing Centre (www.lrz.de). The work is supported by the German Research
Foundation (DFG) research unit FOR5269 "Future methods for studying confined gluons in QCD".
\bibliographystyle{JHEP}
\bibliography{refs} 

\end{document}